\documentclass[%
 reprint,
 superscriptaddress,
 amsmath,amssymb,
 aps,
 prl,
floatfix,
]{revtex4-2}

\usepackage{graphicx}% Include figure files
\usepackage{dcolumn}% Align table columns on decimal point
\usepackage{bm}% bold math
\usepackage{amsfonts}
\usepackage[scr=boondoxo]{mathalfa}
\usepackage{multirow}

\begin{document}

\preprint{APS/123-QED}

\title{Wave-Scattering processes:\\path-integrals designed for the numerical handling of complex geometries}

% Force line breaks with \\

\author{J\'er\'emi Dauchet}
\email{jeremi.dauchet@sigma-clermont.fr}
\affiliation{Université Clermont Auvergne, Clermont Auvergne INP, CNRS, Institut Pascal, F-63000 Clermont-Ferrand, France}
\author{Julien Charon}
\affiliation{ESTACA West Campus, Rue Georges Charpak, 53000, Laval, France}
\author{Laurent Brunel}
\affiliation{PhotonLyx Technology S.L., Santander, Spain}
\author{Christophe Coustet}
\affiliation{Méso-Star, Longages, France}
\author{Stéphane Blanco}
\affiliation{LAPLACE, Université de Toulouse, CNRS, INPT, UPS, Toulouse, France}
\author{Jean-François Cornet}
\affiliation{Université Clermont Auvergne, Clermont Auvergne INP, CNRS, Institut Pascal, F-63000 Clermont-Ferrand, France}
\author{Mouna El Hafi}
\affiliation{Université Fédérale de Toulouse Midi-Pyrénées, Mines Albi, UMR CNRS 5302, Centre RAPSODEE, Campus Jarlard, F-81013 Albi CT Cedex 09, France}
\author{Vincent Eymet}
\affiliation{Méso-Star, Longages, France}
\author{Vincent Forest}
\affiliation{Méso-Star, Longages, France}
\author{Richard Fournier}
\affiliation{LAPLACE, Université de Toulouse, CNRS, INPT, UPS, Toulouse, France}
\author{Fabrice Gros}
\affiliation{Université Clermont Auvergne, Clermont Auvergne INP, CNRS, Institut Pascal, F-63000 Clermont-Ferrand, France}
\author{Benjamin Piaud}
\affiliation{Méso-Star, Longages, France}
\author{Thomas Vourc'h}
\affiliation{Université Clermont Auvergne, Clermont Auvergne INP, CNRS, Institut Pascal, F-63000 Clermont-Ferrand, France}
\date{\today}

\begin{abstract}
Relying on Feynman-Kac path-integral methodology, we present a new statistical perspective on wave single-scattering by complex three-dimensional objects. The approach is implemented on three models - Schiff approximation, Born approximation and rigorous Born series - and usual interpretative difficulties such as the analysis of moments over scatterer distributions (size, orientation, shape...) are addressed. In terms of computational contribution, we show that commonly recognized features of Monte Carlo method with respect to geometric complexity can now be available when solving electromagnetic scattering.
\end{abstract}

\pacs{Valid PACS appear here}% PACS, the Physics and Astronomy
\maketitle

Whether the question is theoretical or numerical, the scattering of waves by objects of complex spatial shape often leads to strong interpretative or computational difficulties, especially when the scatterers are large relative to the wavelength and/or with a high scattering potential \cite{Kahnert2016,VanBladel1991}. 
Furthermore in most application situations, the study of non-spherical scatterers usually requires a statistical description in terms of size, orientation and shape distributions which increases the challenge of obtaining reliable quantifications \cite{Mishchenko2002,Dauchet2015, Wax2010, Liou2016}.

Faced with questions of high complexity in geometrical and phenomenological terms, the choice of alternative representations based on a statistical reformulation can lead to a renewed point of view and produce surprisingly efficient numerical solutions \cite{Villefranque2022,Ibarart2022,DeLaTorre}.
In this perspective, the objective of this letter is to present a novel formulation of the underlying wave physics in probabilistic terms with direct methodological reference to the work of Feynman-Kac~\cite{Botelho}. This requires producing a path space for each of the alternative models we present, and thus formulating the observable as a path integral over this space \cite{Terree2022,Tregan2022,Dauchet2018}. It is then a question of making explicit the quantities of interest directly in the form of an expectation on a stochastic process.

To our knowledge the only prospective work around these ideas has been carried out on the probabilistic reformulation of the electromagnetic model under Schiff approximation for simple shaped scatteres \cite{Charon2016}. In this paper we first extend this work to more complex geometries and then the same approach is applied to other well established path integral formulations of the scattering problem: Born approximation and a rigorous infinite Born series.

The combination of Schiff and Born approximations typically allows the estimation of spectral and angular light scattering properties of photosynthetic microorganisms, or any soft particle, to be used as input to radiative transfer models around issues in the physical chemistry of photobioreactive processes \cite{Dauchet2015}. More generally, our study in complete Born series reveals the difficulties that will be encountered when tackling known sharp problems, such as rigorous solutions for large scattering potentials.

Finally, the proposed formulation offers a new computational perspective by naturally bringing to the forefront Monte Carlo methods (MC) for which the estimation of the expectation of random variables is the most basic theoretical issue.
We show that combining path space sampling with the latest computer graphics tools for intersection calculations is a high-performance solution for obtaining reference calculations for complex scatterer shapes. In addition, we find the commonly recognized characteristics of MC for the simulation of field propagation in optical systems as mentioned in \cite{Prahl2009} : simplicity of treatment of complex boundary conditions, ease of moment estimation on random configurations, etc.

\begin{figure}[h!]
\includegraphics[width=0.45\textwidth]{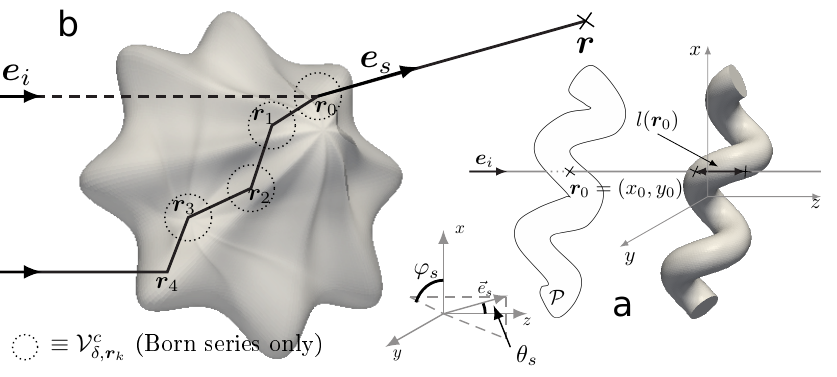}
\caption{Representation of the path spaces associated with the three studied models. The scattered field is estimated at location $\bm{r}$, at distance $r$, in the direction ${\bm e}_s\equiv (\theta_s,\varphi_s)$. {\bf a.} Schiff approximation: paths are straight lines along the incident direction ${\bm e}_i$. {\bf b - dashed black line.} Born approximation: paths are incoming in direction ${\bm e}_i$, interact at one location $\bm{r}_0$ inside the scatterer volume and go out in the scattered direction ${\bm e}_s$. {\bf b - plain black line.} Born series expansion: paths go through multiple interaction locations $\bm{r}_0$, $\bm{r}_1$, $\bm{r}_2$... inside the scatterer.}
\label{Fig_chemins}
\end{figure}

{\bf The scattering problem} is the following: an incident plane wave $\bm{\mathscr E}_{i}$ with wave number $k = \frac{2\pi}{\lambda}$ and propagation direction $\bm{e}_i$ interacts with a scattering potential $U$ defined in a finite region $V$ with complex spacial shape, embedded in an infinite homogeneous and nonabsorbing medium (see Fig.~\ref{Fig_chemins}).
%; in the case of electromagnetic waves the surrounding medium is additionally assumed to be isotropic and nonmagnetic).
The resulting field $\bm{\mathscr E}$ is solution of
\begin{equation}
{\bm\nabla}\times{\bm\nabla}\times\bm{\mathscr E} - k^2\bm{\mathscr E} + U\bm{\mathscr E} = 0
\label{ref:Helmoltz}
\end{equation}
that reduces to Helmoltz equation in case of scalar waves. $\bm{\mathscr E}_{i}$ is solution of the above equation when $U=0$ and the scattered field $\bm{\mathscr E}_{s}$, that is nonzero when $U\neq 0$, is defined as $\bm{\mathscr E}=\bm{\mathscr E}_i+\bm{\mathscr E}_s$. In electromagnetism, the scattering potential is classically defined by the relative refractive index $m$ of the scattering object: $U=k^2(1-m^2)$. The solution of this problem is invariant with respect to contrast $1-m^2$ and size-to-wavelength ratio $x$.

{\bf Path integral formulations} for this problem will be derived from the Volume Integral Equation~\cite{VanBladel1991}
\begin{equation}
\bm{\mathscr E}_s(\bm{r}) = \int_{V} d\bm{r}_0 \,U(\bm{r}_0) \bar{\bar{G}}(\bm{r}, \bm{r}_0) \bm{\mathscr E}(\bm{r}_0)
\label{eq:VIE}
\end{equation}
where $\bar{\bar{\bm{G}}}(\bm{r}, \bm{r}_0) = \left[\bar{\bar{\bm{I}}}+\frac{{\bm\nabla} \otimes{\bm\nabla} }{k^2}	 \right] \frac{e^{-ik \Vert \bm{r}-\bm{r}_0\Vert}}{4\pi \Vert \bm{r}-\bm{r}_0\Vert}$ is the free-space dyadic Green's function.
The objective of the next paragraphs is to express the scattered field as an expectation on a stochastic process:
\begin{equation}
\bm{\mathscr E}_{s}(\bm r) = \mathbb{E}\left[\bm{W}_{\bm{\Gamma}}\right] = \int_{\mathcal{D}_{\bm{\Gamma}}} d\bm{\gamma} \,p_{\bm{\Gamma}}(\bm{\gamma}) \bm{w}(\bm{\gamma})
\label{Eq_Es_chemin}
\end{equation}
where $\mathbb{E}$ is the expectation operator and $W_{\bm{\Gamma}}=\bm{w}(\bm{\Gamma})$ is defined as a function of the random variable $\bm{\Gamma}$ with probability density function (pdf) $p_{\bm{\Gamma}}$ over $\mathcal{D}_{\bm{\Gamma}}$. We will focus on formulations in which $\mathcal{D}_{\bm{\Gamma}}$ is a path space and $W_{\bm{\Gamma}}$ is the contribution to $\bm{\mathscr E}_{s}$ of the random path $\bm{\Gamma}$. Computationally speaking, it leads to MC algorithms sampling $N$ paths $\gamma_i$ to finally estimate their average contribution $\bm{\mathscr E}_{s}\simeq1/N\sum_{i=1}^N \bm{w}(\gamma_i)$, with statistical error provided by the standard deviation of the contributions $\bm{w}(\gamma_i)$. The following paragraphs aim at producing random paths $\bm{\Gamma}$ that can be efficiently sampled whatever the scatterer geometry, using state-of-the-art computer graphics tools.

{\bf Schiff approximation}~\cite{Schiff1956} is an eikonal-like approximation based on Eq.~\ref{eq:VIE} that gives the scattered field in the far-field region for large soft-scatterers ($x \gg 1$ and $\vert m-1\vert \ll 1$). In~\cite{Charon2016}, it is reformulated as an expectation which leads to, with the notations of Eq.~\ref{Eq_Es_chemin}:
\begin{equation}
\begin{split}
\bm{W}_\Gamma =\mathcal{P}\frac{ik}{2\pi} \frac{e^{-ik r}}{r}&
e^{ik\theta_s ({X}_0\cos \varphi_s+ {Y}_0\sin \varphi_s)} \\ &\times 
 \left[1-e^{-ik(m-1) l(\bm{R}_0)} \right] 
\end{split}
\label{Eq_Es_Schiff_W}
\end{equation}
where $\bm{R}_0 =(X_0,Y_0)$ is a random location uniformly distributed on the scatterer's projected surface $\mathcal{P}$ seen from a given incident direction ${\bm e}_i$, and $l(\bm{R}_0)$ is the crossing length of the straight path starting at $\bm{R}_0$ in the direction ${\bm e}_i$ (see a realization in Fig.~\ref{Fig_chemins}.a). The trial for the path $\bm{\Gamma}$ and its contribution $W_{\bm{\Gamma}}$ is the following:
1. a location $\bm{r}_0$ is uniformly sampled over $\mathcal{P}$, 2. the path $\gamma_i$ is traced from $\bm{r}_0$ in direction ${\bm e}_i$, 3. the path crossing length $l(\bm{r}_0)$ is retrieved and the contribution is calculated according to Eq.~\ref{Eq_Es_Schiff_W} with $\bm{R}_0=\bm{r}_0$.

{\bf Born approximation}, that is similar to the Rayleigh-Gans-Debye approximation \cite{Bohren1983}, is valid for small soft-scatterers ($x \ll 1$ and $\vert m^2-1\vert \ll 1$). For observation points $\bm r$ in the far-field region, it assumes that the field inside the scatterer is equal to the incident field, {\it i.e.} $\bm{\mathscr E}(\bm{r}_0)=\bm{\mathscr E}_{i}(\bm{r}_0)$ in Eq.~\ref{eq:VIE}. Following the methodology presented in~\cite{Charon2016}, this equation is multiplied and divided by the pdf $p_{\bm{R}_0}(\bm{r}_0)$ of a random location $\bm{R}_0$ defined over the scatterer's volume $V$ (at this stage, we don't need to specify this pdf except that it is non-zero for every location $\bm{r}_0 \in V$ ; take a uniform distribution for example):
\begin{equation}
\bm{W}_\Gamma =\frac{U(\bm{R}_0)  \bar{\bar{\bm{G}}}(\bm{r}, \bm{R}_0) \bm{\mathscr E}_{i}(\bm{R}_0)}{p_{\bm{R}_0}(\bm{R}_0)}
\label{Eq_BA_weight_E}
\end{equation}
Here the path $\bm{\Gamma}(\bm{r},\bm{R}_0)$ comes from direction $\bm{e}_i$, interacts at location $\bm{R}_0$ within $V$ and leaves the scatterer in direction $\bm{e}_s$ until it reaches $\bm{r}$ (see  a realization in Fig.~\ref{Fig_chemins}.b). The trial corresponding to Eq.~\ref{Eq_BA_weight_E} is: 1. a location $\bm{r}_0\in V$ is sampled according to $p_{\bm{R}_0}$, 2. the path $\gamma_i$ is traced, 3. its contribution is calculated according to Eq.~\ref{Eq_BA_weight_E} with $\bm{R}_0=\bm{r}_0$.

\begin{figure}[h!]
\includegraphics[width=\columnwidth]{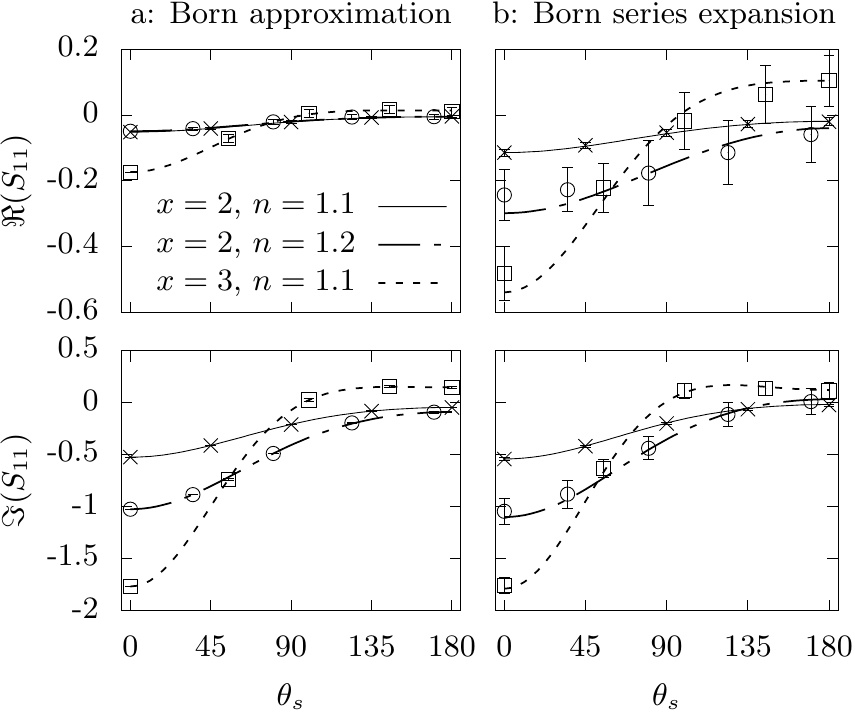}
\caption{Results for Born approximation (a - left panels) and Born series expansion (b - right panels) in the far field region, for spheres with different size parameters $x$ and refractive indices $m=n-i\,0.01$. Real part (upper panels) and imaginary part (lower panels) of the first element in the scattering matrix $\bar{\bar{S}}$ are plotted as function of the scattering angle $\theta_s$ (see Fig.~\ref{Fig_chemins}), using the definition in~\citep{Bohren1983}: $\bm{\mathscr E}_{s}=\frac{e^{ikr}}{-ikr}\bar{\bar{S}}\,\bm{\mathscr E}_{i}$. Analytical reference solutions from~\citep{Bohren1983} are plotted with lines. MC estimates obtained for $10^6$ paths are plotted with points and errorbars indicating the $99\%$ confidence interval. The algorithm implemented to produce these results uses an optimization consisting in summing, along the same path $\Gamma|M$, the contributions $W_{\Gamma|N=0}$, $W_{\Gamma|N=1}$, $\cdots$ $W_{\Gamma|N=M}$ of several orders $N=1,2,\cdots, M$ in the Born series, where $M$ is a random truncation order (see details in SM).
The first location $\bm{R}_0$ is uniformly sampled - the other sampling distributions are given in SM - and the radius of the exclusion volume is $\delta=\frac{x^2}{k}\frac{n^2-1}{n^2+2}$. Computation time is about 1 s for Born approximation and 3 s for Born series expansion on a Intel Core i7-3720QM@2.60GHz CPU laptop.
}
\label{fig:Mie}
\end{figure}

{\bf Born series expansion} gives a reference solution. In comparison with Born approximation, the internal field $\bm{\mathscr E}(\bm{r}_0)$ is no more approximated, but obtained by applying Eq.~\ref{eq:VIE} for locations inside ${\cal V}$. In this case, the strong singularity at ${\bf r}={\bf r}_0$ can be handled with the Cauchy principal value for spherical exclusion volume $V_{\delta,\bm{r}}$ of radius $\delta$ centered at ${\bf r}$~\cite{VanBladel1991}. Eq.~\ref{eq:VIE} becomes
\begin{equation}
\bm{\mathscr E}(\bm{r}_0) = \eta(\bm{r}_0)\bm{\mathscr{E}}_{i}(\bm{r}_0) + (\bm{\bar{\bar{\mathcal L}}}^1\bm{\mathscr E})(\bm{r}_0)
\label{ref:VIE-Vexclu}
\end{equation}
with $\eta(\bm{r}) =\frac{3}{m^2(\bm{r})+2}$ and the integral operator
\begin{equation*}
(\bm{\bar{\bar{\mathcal L}}}^n\bm{f})(\bm{r}_0) = \lim_{\delta\rightarrow 0} \int_{V^c_{\delta,\bm{r}_0}}\hspace{-6mm} d\bm{r}_1\cdots\int_{V^c_{\delta,\bm{r}_{n-1}}}\hspace{-6mm} d\bm{r}_n  \prod_{j=1}^n \bar{\bar{A}}(\bm{r}_{j-1}, \bm{r}_j)\,\bm{f}(\bm{r}_n)
\end{equation*}
where $\bar{\bar{A}}(\bm{r}_{j-1}, \bm{r}_j)=U(\bm{r}_j)\eta(\bm{r}_{j-1}) \bar{\bar{G}}(\bm{r}_{j-1}, \bm{r}_j)$ and $V^c_{\delta,\bm{r}}=V \setminus V_{\delta,\bm{r}}$ is the set of locations in the scatterer volume but not in the exclusion volume (it is the complement of $V_{\delta}(\bm{r})$ in $V$). Numerically, we classically ensure that the value of $\delta$ is small enough (see Fig.~\ref{fig:Mie}). 

The Born series expansion of the internal field is obtained by successive substitution of Eq.~\ref{ref:VIE-Vexclu} into itself: $\bm {\mathscr  E}(\bm{r}_0) = \sum_{n=0}^{+\infty}(\bm{\bar{\bar{\mathcal L}}}^n \eta\,\bm {\mathscr  E}_i)(\bm{r}_0)$. This expression is exact only for values of $m$ and $x$ within the radius of convergence of the series~\cite{Kilgore2017}. Reformulation as an expectation requires two steps: first, the integral operator $\bm{\bar{\bar{\mathcal L}}}^n$ is reformulated using $n$ random locations $\bm{R}_{j=1,2,...,n}$ with distribution $p_{\bm{R}_j}(\bm{r}_j)$ and second, the infinite sum in Born series is reformulated by introducing a discrete random variable $N$ - a random order in the series - with probability distribution $p_N$ (each term in the sum is multiplied and divided by $p_N(n)$, with $\sum_{n=0}^{+\infty}p_N(n)=1$). % : 
Finally, the incident field in Eq.~\ref{Eq_BA_weight_E} is replaced by this expression of the internal field, to obtain Eq.~\ref{Eq_Es_chemin} with:
\begin{equation}
\bm{W}_\Gamma=\frac{U(\bm{R}_0)  \bar{\bar{\bm{G}}}(\bm{r}, \bm{R}_0)}{p_{\bm{R}_0}(\bm{R}_0)}\prod_{j=1}^N \frac{\bar{\bar{A}}(\bm{R}_{j-1}, \bm{R}_j)}{p_{\bm{R}_j}(\bm{R}_j)}\,\frac{\eta(\bm{R}_N)\bm {\mathscr  E}_i(\bm{R}_N)}{p_N(N)}
\label{Eq_B_weight_E}
\end{equation}
Intermediate steps leading to this result are provided in the Supplemental Material (SM).
The path $\bm{\Gamma}(\bm{r},\bm{R}_0,\bm{R}_1,\cdots,\bm{R}_N)$ comes from direction $\bm{e}_i$, interacts at several locations $\bm{R}_0,\bm{R}_1,\bm{R}_2\cdots$ within $V$ and leaves the scatterer in direction $\bm{e}_s$ until it reaches $\bm{r}$ (see Fig.~\ref{Fig_chemins}.b). The trial corresponding to Eq.~\ref{Eq_B_weight_E} is: 1. a location $\bm{r}_0\in V$ is sampled according to $p_{\bm{R}_0}$, 2. the number $n$ of locations for the current path $\gamma_i$ is sampled according to $p_N$, 3. $\gamma_i$ is traced by successively sampling the $n$ locations $\bm{r}_{j=1,2\cdots,n}$ according to their respective distributions $p_{\bm{R}_j}$ in $V^c_{\delta,\bm{r}_{j-1}}$, 3. the contribution is calculated according to Eq.~\ref{Eq_B_weight_E} with $N=n$ and $\bm{R}_j=\bm{r}_j$ (if $n=0$, the contribution is $U(\bm{r}_0)  \bar{\bar{\bm{G}}}(\bm{r}, \bm{r}_0)\eta(\bm{r}_0)\bm {\mathscr  E}_i(\bm{r}_0)/p_{\bm{R}_0}(\bm{r}_0)$). 

{\bf Numerical validation} concerning Schiff approximation have already been presented in~\cite{Charon2016}; only computation times for complex geometries will be reported in Table~\ref{tab:StarSchiff}. The results obtained for Born approximation and Born series expansion are presented in Fig.~\ref{fig:Mie} and cross validated with reference analytical solutions for spheres.
However, no simplification related to spherical shape is used here.
The geometry of the scatterer only affects path sampling: 
locations $\bm{R}_j$ are here sampled within a sphere; more complex scatterers would only require to sample a more complex volume, as discussed later. 

Note that the estimator for Born approximation converges quickly and no difficulty is recorded (see Fig.~\ref{fig:Mie}.a).
For Born series expansion (see Fig.~\ref{fig:Mie}.b), by contrast, we record convergence issues when increasing refractive index (compare error bars for crosses and circles) and size parameter (compare crosses and squares).
This was expected since numerical solution of Maxwell's equations is known to be difficult in that case. 
Usual deterministic numerical methods are mainly limited by current computer memory size and/or floating point accuracy~\citep{Kahnert2016}. Here, limitations are totally different. 
We only observe convergence issues that should be subsequently addressed using well established integral reformulation approaches such as zero-variance principle to optimize the sampling distributions $p_{\bm{R}_j}$ (see the current $p_{\bm{R}_j}$ choices in SM).

{\bf Average cross section $\sigma(\bm{e}_s)=r^2\,\bm{\mathscr  E}_s\cdot\bm{\mathscr  E}_s^*$ of an ensemble of scatterers} is often a targeted quantity when solving scattering problems~\cite{Dauchet2015,Wax2010,Liou2016,Mishchenko2002,Schiff1956}. But injecting Eq.~\ref{Eq_Es_chemin} in this expression does not lead to an expectation, due to Poynting like nonlinearity with respect to $\bm{W}_{\bm{\Gamma}}$: $\sigma(\bm{e}_s)=r^2\,\mathbb{E}\left[\bm{W}_{\bm{\Gamma}}\right]\cdot\mathbb{E}\left[\bm{W}_{\bm{\Gamma}}^*\right]\neq \mathbb{E}\left[r^2\,\bm{W}_{\bm{\Gamma}}\cdot\bm{W}_{\bm{\Gamma}}^*\right]$. Following the methodology presented in~\cite{Dauchet2018,Charon2016}, $\sigma(\bm{e}_s)$ is reformulated as the expectation on a stochastic process in the squared path-space $\mathcal{D}_{\bm{\Gamma}}^2$, by using two independent random path variables $\bm{\Gamma}_1$ and $\bm{\Gamma}_2$ identically distributed as $\bm{\Gamma}$:
\begin{equation}
\sigma(\bm{e}_s) = \mathbb{E}\left[r^2\bm{W}_{\bm{\Gamma}_1}\cdot\bm{W}_{\bm{\Gamma}_2}^*\right]
\label{Eq:section}
\end{equation}
Averaging over scatterer distributions is now straightforward, leading to the following trial: 1. shape, orientation, size, value of the scattering potential $U$ are sampled, 2. two paths $\bm{\gamma}_1$ and $\bm{\gamma}_2$ are sampled (see the procedures in previous paragraphs) and 3. the contribution $r\,\bm{w}_{\bm{\gamma}_1}\cdot r\,\bm{w}_{\bm{\gamma}_2}^*$ is computed. A direct consequence is that MC calculation times are weakly sensitive to refinement of the statistics of geometry, since path space and geometric configuration space are simultaneously covered~\cite{Weitz2016} (see Table~\ref{tab:StarSchiff}).

\begin{table}[ht]
\begin{ruledtabular}
\begin{tabular}{cp{3.75cm}ccc}
\multirow{11}{*}{\includegraphics[width=2.2cm,height=4.75cm]{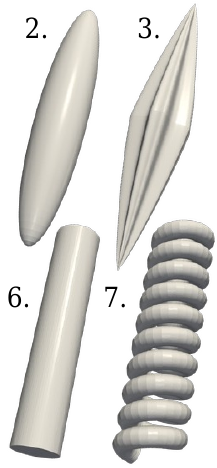}} & Geometry & $\frac{N}{10^5}$ & $t\,$(s) & $m\,$({\scriptsize Mo}) \\
\vspace{-2mm}\\
  & 1. Monodisperse sphere: & 0.9	&0.04	&22 \\
  & 2. Monodisperse ellipsoid: & 4	&0.20	&87 \\
  & 3. Monodisp. supershape: & 8	&0.35	&92 \\
  & 4. Polydisp. supershape: & 15	&0.55	&93 \\
  & 5. Distributed supershape: & 15	&18.1	&147 \\
\vspace{-2mm}\\
  & 6. Monodisperse cylinder: & 2	&0.15	&18 \\
  & 7. Monodisperse helix: & 4	&0.30	&56 \\
  & 8. Polydisperse helix: & 6	&0.58	&56 \\
  & 9. Distributed helix: & 6	&4.20	&88 \\
  & 10. One-ninth mixture of &  & & \\
  & the above shapes: & 11	&12.2	&193 \\
\end{tabular}
\end{ruledtabular}
\caption{Computation of total cross-sections (extinction, scattering, absorption) and differential cross-section at $\theta_s=1^\circ$ with schiff software~\cite{MesoStar} (Schiff approximation for soft particles). Number of samples $N$, calculation times $t$ and peak memory usage $m$ required to achieve standard error $<1\%$ for various shapes, with the same laptop used in Fig.~\ref{fig:Mie}. "Monodisperse" indicates a unique geometry, "Polydisperse" a log normal size distribution with $\ln(\sigma)=0.18$ and "Distributed" a distribution of several parameters in the shape parametric equation. Orientation is isotropically distributed.
Properties are representative of photosynthetic microorganisms: $m=1.1\,-i\,5\,10^{-3}$, $\lambda=400\,nm$, volume-equivalent sphere radius $r_{eq}= 6\,\mu m$, {\it i.e.} $x=\frac{2\pi r_{eq}}{\lambda}\simeq 94$ (on average when size is distributed) and aspect ratio 1/5 (on average when shape is distributed). Shapes and commands used to produce those results are provided in the SM.}
\label{tab:StarSchiff}
\end{table}

{\bf Implementation for complex geometries} is commonly recognized to be quite simple with MC approaches. Indeed, using open access libraries for ray tracing developed by the computer graphics community under the solicitation of the cinema industries \cite{Pharr2016}, we are able to easily implement the path sampling procedures presented in this paper for any geometry specified by its bounding surface. However, sampling the geometric data during MC calculation, as required here, is not straightforward with available tools because they usually aim at generating images from fixed scenes (and animations are constructed as sequences of such images). For that reason, most ray-tracing acceleration structures have been developed for fixed geometric data and generating such a structure at each MC sample would be highly inefficient. We therefore developed a specific approach in collaboration with computer graphics experts \cite{MesoStar}. The geometry is specified by statistical distributions of parameters and based on~\citep{Weitz2016}, several paths are traced for each sampled shape. We fully implemented this approach for Schiff's approximation and the result is a free and open-source software whose features with respect to geometric complexity are presented in Table~\ref{tab:StarSchiff}. An outstanding feature of this programming approach is the orthogonality between the path tracking algorithm and the representation of surfacic data~\cite{Novak2014, Kutz2017}. As a result, simulating any complex-shaped scatterer is in practice as simple as simulating a sphere and calculation times are weakly sensitive to the shape complexity (compare cases 2, 3, 6 and 7 in Table~\ref{tab:StarSchiff}). Memory requirements also remain quite stable and low compared to deterministic methods using discretization ({\it e.g.} method-of-moments) since MC has the ability to directly evaluate targeted quantities without having to generate mesh and compute intermediate fields over it.  Furthermore, thanks to the work on path-integral formulation presented above~\cite{Charon2016,Weitz2016,Dauchet2018} the MC samples required to solve the scattering problem are also used to cover the statistics of geometry simultaneously. As a result, the number of samples and computation time are only multiplied by 5 when accounting for orientation distribution (compare cases 2 and 6 with case 1), by 2 when further adding size distribution (compare cases 4 and 8 with 3 and 7 respectively) and evaluating the properties of a mixture of 9 particle types in case 10 is easier than simulating the most demanding particle type alone (here case 5). In our examples, the number of samples required to simulate size distribution is sufficient to also cover aspect ratio and shape distribution ($N$ is the same in cases 4 and 5, and in cases 8 and 9) but the computation time is increased. This additional time corresponds to the generation of the geometric data in the case of distributed objects, while we avoid such geometry sampling in the case of size distribution thanks to a scaling of the wavelength that preserves size-to-wavelength ratio $x$. Finally, this software also takes advantage of the opportunity to estimate several quantities simultaneously with the same paths sample~\citep{Charon2016,DeLaTorre}. As a result, evaluating 40 wavelengths to construct a spectrum, as in~\cite{Dauchet2015}, only multiplies calculation time by 4 - instead of 40 with deterministic methods - when compared to calculation for one single wavelength in Table~\ref{tab:StarSchiff}. Overall, we are now able to produce spectral and angular radiative properties of helical shaped microalgae {\it Arthrospira platensis} in 20 minutes, while it was requiring months with a straight cylinder model using deterministic integration methods~\cite{Dauchet2015}.

{\bf To conclude}, we have presented a new statistical formulation for wave scattering that brings original path-integral representations and high-performance numerical solutions. High-contrast and large scatterers remains challenging, as with deterministic approaches, and will require joint efforts in path-integral formulation for hyperbolic differential equation, as well as sampling strategies for the resulting processes. Although our examples have focused on electromagnetic scattering perspectives, let us emphasize that the three models illustrated here have been developed in the field of quantum mechanics.

\begin{acknowledgments}
This letter synthesizes the PhD work~\cite{charonThese} sponsored by grants ANR-10-LABX-0016 (Labex IMobS3), ANR-16-IDEX-0001 (IDEX-ISITE CAP 20-25) and ANR-10-LABX-22–01 (Labex SOLSTICE).
\end{acknowledgments}

%\appendix
%
%\section{Appendixes}
%
\nocite{*}

\bibliography{apssamp}% Produces the bibliography via BibTeX.

\end{document}